\documentclass[prd,aps,preprint,tightenlines]{revtex4}
\usepackage{epsfig}
\begin{document}
\voffset=0.5cm
\preprint{UMD PP\#00-080}
\preprint{DOE/ER/40762-208}
\title{ONE-LOOP FACTORIZATION OF THE NUCLEON\\
$g_2$-STRUCTURE FUNCTION IN THE NON-SINGLET CASE}

\author{Xiangdong Ji}
\email{xji@physics.umd.edu}
\affiliation{Department of Physics, 
University of Maryland, 
College Park, Maryland 20742 }
\author{Wei Lu}
\email{weiluyao@hotmail.com}
\affiliation{Department of Physics, 
University of Maryland, 
College Park, Maryland 20742 }
\author{Jonathan Osborne}
\email{jao@physics.umd.edu}  
\affiliation{Department of Physics, 
University of Maryland, 
College Park, Maryland 20742 }
\author{Xiaotong Song}
\email{xs3e@virginia.edu}
\affiliation{Institute for Nuclear and Particle Physics, Department 
of Physics, University of Virginia, Charlottesville, Virginia 22904}
\vspace{0.2in}
\date{\today}

\begin{abstract}
We consider the one-loop
factorization of the simplest twist-three process: 
inclusive deep-inelastic scattering of longitudinally-polarized 
leptons on a transversely-polarized nucleon
target. By studying the Compton amplitudes
for certain quark and gluon states at one loop, we find the 
coefficient functions for the non-singlet twist-three 
distributions in the factorization formula of
$g_2(x_B,Q^2)$. The result marks the first
step towards a next-to-leading order (NLO) formalism 
for this transverse-spin-dependent structure function
of the nucleon. 
\end{abstract}
\maketitle

Deep-inelastic scattering (DIS) of leptons on the nucleon 
is a time-honored example of the success
of perturbative quantum chromodynamics (PQCD)\cite{mueller}. 
The factorization formulae for the leading structure 
functions $F_1(x_B, Q^2)$ and $g_1(x_B, Q^2)$, 
augmented by the Dokshitzer-Gribov-Lipatov-Altarelli-Parisi (DGLAP) 
evolution equations for parton distributions
\cite{dglap}, can describe the available DIS data collected over the last 
30 years exceedingly 
well. Although the same formalism is believed
to work for the so-called higher-twist structure 
functions \cite{sterman}, e.g. $g_2(x_B, Q^2)$ and $F_L(x_B, Q^2)$,
which contribute
to physical observables down by powers of the hard 
momentum $Q$, there are few detailed 
studies of them in the literature beyond the tree level. 
The QCD radiative corrections to 
$g_2(x_B,Q^2)$ need be investigated as accurate
data have recently been taken \cite{e155x} and more data will 
be available in the future \cite{jlab}.
 
In this paper, we report a one-loop study
of inclusive deep-inelastic scattering of  
longitudinally-polarized leptons (e.g. electrons)
on a transversely-polarized nucleon target \cite{hey}. 
The subject was first investigated in the context of single parton 
scattering in Ref. \cite{kodaira}, and studies
along the same line have continued 
in the literature \cite{quarkonly}.  
However, due to the subtlety of the twist-three
process \cite{sv,bkl,other}, those results are
sensitive to the treatment of quark masses and
are incomplete in the context of QCD fatorization.  
Indeed, even at tree level one must go beyond 
the single quark process to derive the 
correct $g_2(x_B, Q^2)$ expression in terms of the parton
distributions \cite{et,jj,ji}. When loop corrections
are included, one needs a general strategy to 
systematically calculate their contribution
to higher twist processes.

For a transversely polarized nucleon of four-momentum $P^\mu$
and polarization vector $S_\perp^\mu$, the hadron tensor 
$W^{\mu\nu}  = {1\over 4\pi}\int e^{iq\cdot \xi}
  \langle PS|[J^\mu(\xi),J^\nu(0)]PS\rangle$
can be expressed as 
\begin{equation}
    W^{\mu\nu} = - i\epsilon^{\mu\nu\alpha\beta}
   q_\alpha S_{\perp\beta}{1\over \nu}\left(g_1(x_B, Q^2) 
   + g_2(x_B, Q^2)\right) \ , 
\end{equation}
where $q^\mu$ is the photon four-momentum, $Q^2=-q^2$, 
$\nu=q\cdot P$, $x_B = Q^2/(2\nu)$, 
and $\epsilon^{0123}=+1$. $J^\mu$ is the electromagnetic
current of the quarks.
Thus it is the combination $g_T(x_B, Q^2) \equiv g_1(x_B, Q^2)
  + g_2(x_B, Q^2)$ that naturally appears in the
$1/Q$-suppressed transverse polarization asymmetry. 
In this paper, we concentrate only on the non-singlet part
of $g_T(x_B, Q^2)$; the singlet case is left to 
a separate publication. We seek a factorization formula 
for $g_T(x_B,Q^2)$ to one-loop order 
in terms of the perturbative coefficient functions $C_i(x,y)$ and 
the generalized parton distributions (correlations) $K_i(x,y)$ 
with two light-cone (or Feynman) variables $x$ and $y$.
The starting point is one-loop forward virtual-photon 
Compton scattering off a few ``on-shell" quark and gluon states. 
>From the scattering amplitudes, we examine the validity of 
infrared factorization and extract the one-loop
coefficient functions $C_i(x,y)$. 

We begin by outlining a general 
approach to the factorization of higher-twist observables, 
generalizing the method of Ref. \cite{efp} for the 
twist-two structure functions. For deep-inelastic scattering, 
it is convenient to consider first the forward Compton amplitude 
\begin{equation}
    T^{\mu\nu} = i\int d^4\xi e^{iq\cdot\xi}
     \langle PS|{\rm T} J^\mu(\xi) J^\nu(0)|PS\rangle \ . 
\end{equation}
The spin-dependent part of the 
Compton amplitude (antisymmetric in $\mu$ and $\nu$) 
defines two invariant amplitudes $S_{1,2}(x_B, Q^2)$, 
\begin{equation}
    T^{\mu\nu} = -i\epsilon^{\mu\nu\alpha\beta}q_\alpha
   \left({S_1(x_B, Q^2)\over \nu}S_\beta +
    {S_2(x_B, Q^2)\over \nu^2}\left(\nu S_\beta
   - q\cdot S P_\beta \right)\right) \ . 
\end{equation}
According to the optical 
theorem, the imaginary part of the $S_{1,2}$ amplitudes (divided by
$2\pi$) is just the nucleon spin structure
functions $g_{1,2}$. Hence, factorization 
of $S_T(x_B, Q^2) = S_1(x_B, Q^2)+
S_2(x_B, Q^2)$ naturally leads to  
factorization of the structure function $g_T(x_B, Q^2)$.
The former has a straightforward 
Feynman-Dyson perturbative expansion, making it
easily handled.

According to Ref. \cite{efp}, 
the nucleon Compton amplitude $T^{\mu\nu}$
can be expressed as a sum of terms that are
convolutions of quark-gluon Compton
scattering amplitudes $M_i^{\mu\nu}$ and 
the bare quark-gluon correlation functions $\Gamma_{iB}$
in the nucleon, as shown schematically in Fig. 1,   
\begin{equation}
      T^{\mu\nu} = \sum_i M^{\mu\nu}_i \otimes \Gamma_{iB} \ .
\label{conv}
\end{equation}
Implicitly involved in the convolution are integration
over the intermediate quark-gluon four-momenta $k_j$
and summation over the spin and color indices.
$M^{\mu\nu}_i$, without external quark-gluon 
legs and self-energies, is the 
sum of a complete set of Feynman diagrams for 
Compton scattering. These diagrams are calculated
in unrenormalized perturbation theory in the sense that 
all parameters in the expansion
are bare. We use dimensional regularization 
to regularize both infrared and ultraviolet divergences in 
the diagrams. 

\begin{figure}[t]
\begin{center}
\epsfig{file=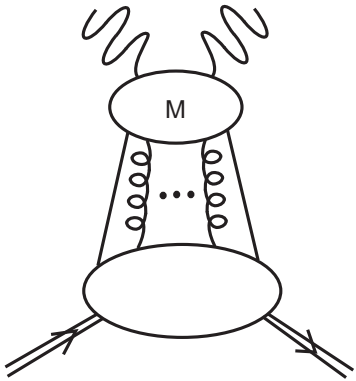,clip=,height=5cm,angle=0}
\end{center}
\caption{Schematic diagram for Compton scattering in terms 
of quark and gluon scattering amplitudes and their correlation functions 
in the nucleon.}
\label{fig:1}
\end{figure}   

The pair of light-cone four-vectors 
$p^\mu =\Lambda(1,0,0,1)/\sqrt{2}$ and $n^\mu=(1,0,0,-1)/(\sqrt{2}\Lambda)$  
define a collinear basis in which the nucleon 
and photon momenta are written $P^\mu=p^\mu+M^2 n^\mu/2$ and
$q^\mu=-x_Bp^\mu+ \nu n^\nu$, respectively, where $\Lambda$
is an arbitrary dimensionful parameter.
Intermediate quark and gluon momenta 
can also be expressed in this basis:
\begin{equation}
     k_j^\mu = (k_j\cdot n) p^\mu + (k_j\cdot p) n^\mu + k_j^\perp \ , 
\end{equation}
and the full result for
$M^{\mu\nu}_i$ can be expanded about $k_j^\mu = (k_j\cdot n) p^\mu$.  
The leading term in this expansion can be interpreted 
as scattering of collinear partons with Feynman momentum fractions 
$x_j = k_j\cdot n$. Because
the partons are on-shell, $M^{\mu\nu}_i$ can be 
viewed as the parton scattering S-matrix element. In principle, 
one must multiply by parton wave function renormalization 
factors to get the proper S-matrix element. 
However, in dimensional regularization, the absence
of a physical scale at the massless poles 
of the quark and gluon propagators reduces these
contributions to unity.
Subleading terms in the expansion of $M^{\mu\nu}_i$
are parton scattering S-matrix elements with insertions
of certain vertices associated with
powers of $k_j^\perp$ and $k^-_j(\equiv p\cdot k_j)$. 
For instance, with one power of quark 
momentum $k_j^\perp$, the subleading term 
is calculated with one insertion of the vector
vertex $i\gamma^\alpha_\perp$ to one of the quark propagators.
Because S-matrix elements and their relatives are 
gauge invariant, one may choose any gauge for the internal
gluon propagators in $M^{\mu\nu}_i$. We use 
Feynman's choice in our calculation.

After the collinear expansion and integration over the quark-gluon
four momenta, the convolution of Eq. (\ref{conv}) involves
integrating over the parton Feynman variables $x_j = (k_j\cdot n)$. 
The correlation functions $\Gamma_{iB}$ are matrix elements 
of gauge-invariant nonlocal quark and gluon 
operators in one-to-one correspondence with the external parton states 
in $M^{\mu\nu}_i$. In particular, when $M^{\mu\nu}_i$ contains 
momentum-related vertex insertions, the quark and/or gluon fields
in $\Gamma_{iB}$ appear with partial derivatives. The collinear expansion
results in all QCD fields separated in spacetime along 
the light-cone direction $n^\mu$. As in the 
leading-twist case, expanding the gluon polarization indices
and summing over all contributions from longitudinally-polarized
gluons generates straightline gauge-links which connect 
fields at separate spacetime points. To simplify 
the derivation of a factorization formula, we use 
the light-cone gauge ($A\cdot n=A^+=0$), or link-free, 
expression for $\Gamma_{iB}$. 
This gauge choice allows one to focus on the physical partons 
without worrying about the effects of the longitudinally polarized
gluons. The gauge-invariant form of the factorization 
is recovered simply by imposing gauge invariance on the
final form of parton correlations.

As we have argued above, $M_i^{\mu\nu}$ is
ultraviolet finite because the on-shell 
wave function renormalization is trivial in dimensional
regularization. Nevertheless, due to the massless
on-shell external states, $M_i^{\mu\nu}$ has
infrared divergences showing up as $1/\epsilon$ poles. These 
divergences may be factorized in the perturbative sense  
\begin{equation}
      M^{\mu\nu}_i = C^{\mu\nu}_i \otimes P_i \ , 
\end{equation}
where $C_i^{\mu\nu}$ is the finite coefficient function 
and $P_i$ contains only the $1/\epsilon$ poles.
On the other hand, the infrared-finite quantities $\Gamma_{iB}$ 
contain ultraviolet divergences which also show 
up as $1/\epsilon$ poles. When the infrared poles 
in $P_i$ cancel all the ultraviolet poles 
in $\Gamma_{iB}$, $T^{\mu\nu}$ is said to be 
factorizable.  The product $P_i\Gamma_{iB}$
defines the renormalized parton correlation 
functions $\Gamma_i$. The final factorization formula
for the Compton amplitude is then
\begin{equation}
      T^{\mu\nu} = \sum_i C^{\mu\nu}_i\otimes \Gamma_i\  , 
\end{equation}
where $C^{\mu\nu}_i$ is a well-defined 
perturbation series in $\alpha_s$
and $\Gamma_i$ is a finite nonperturbative distribution.

\begin{figure}[t]
\begin{center}
\epsfig{file=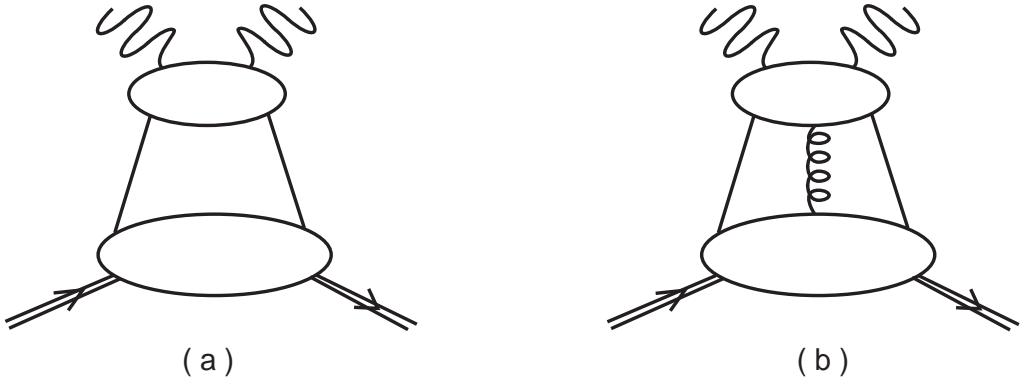,clip=,height=5cm,angle=0}
\end{center}
\caption{Parton intermediate states appearing in the twist-three
Compton amplitude.}
\label{fig:60_1}
\end{figure}

We apply the above discussion to Compton scattering on a
transversely polarized nucleon. As shown in Fig. 2, 
the non-singlet twist-three process
involves two possible intermediate parton states in the
light-cone gauge. First is the two-quark state 
with the transverse momentum flowing 
through $M_{qq}^{\mu\nu}$, as shown in Fig. 2a. After 
Taylor-expansion, we keep only the contribution with exactly 
one insertion of the $i\gamma^\alpha_\perp$ vertex. 
The corresponding 
correlation function is
\begin{eqnarray}
    \Gamma_{qqB}^\alpha(x,y) &=& \int {d\lambda \over 2\pi}
      {d\mu\over 2 \pi} e^{i\lambda x}
      e^{i\mu(y-x)}\langle PS|\bar \psi(0)
     i\partial^\alpha_\perp \psi(\lambda n) 
       |PS\rangle \nonumber \\
     & = & \delta(x-y) \int {d\lambda\over
     2\pi}e^{i\lambda x}\langle PS|\bar \psi(0)
    i\partial^\alpha_\perp \psi(\lambda n) |PS\rangle \ , 
\label{vanish}
\end{eqnarray}
where the Dirac and color indices on quark fields are open
and $\alpha$ is perpendicular to $p^\mu$
and $n^\mu$. The second intermediate parton state 
involves two quark lines and one gluon line, as shown in Fig. 2b. 
If we use the Feynman rule $it^a\gamma^\alpha_\perp$ for the 
gluon attachment to $M^{\mu\nu}_{qgq}$, the corresponding correlation 
function is 
\begin{equation}
    \Gamma_{qgqB}^\alpha(x,y) = \int {d\lambda \over 2\pi}
      {d\mu\over 2 \pi} e^{i\lambda x}
      e^{i\mu(y-x)}\langle PS|\bar \psi(0)
     (-g_B)A^\alpha_\perp(\mu n) \psi(\lambda n) 
       |PS\rangle \ . 
\end{equation}
We remind the
reader that all fields and the couplings here are bare.
To decouple the spin and color indices in $M^{\mu\nu}_i$
and $\Gamma_{iB}$, we write explicitly
\begin{eqnarray}
    \Gamma_{qqB}^{\alpha i j}(x, y) &=& 
       \left(S_\perp^\alpha \gamma_5\not\! p \Gamma_{1(qq)B}(x,y)
       + iT^\alpha_\perp \not\! p \Gamma_{2(qq)B}(x, y)\right)
     {\delta_{ij}\over N_c} \ , \nonumber \\
    \Gamma_{qgqB}^{\alpha i j a}(x, y) &=&\left(S_\perp^\alpha 
   \gamma_5\not\! p\Gamma_{1(qgq)B}(x,y)
       + iT^\alpha_\perp \not\! p \Gamma_{2(qgq)B}(x, y)\right)
     {t^a_{ij}\over N_c C_F} \ , 
\label{decom}
\end{eqnarray}
where $N_c$ is the number of colors,  $C_F=(N_c^2-1)/(2N_c)$, 
and $T^\alpha = \epsilon^{\alpha\beta\gamma\delta} S_{\perp \beta}
p_\gamma n_\delta$. In dimensional regularization, $\gamma_5$ must be
defined explicitly.  Here, we follow
t' Hooft and Veltman's convention.  

With appropriate insertions of light-cone gauge links,
which can be generated by summing over intermediate states with additional
longitudinally-polarized gluons, we define the following 
gauge-invariant parton correlations \cite{et,ji}:
\begin{equation}
     K_{iB}(x,y) = \Gamma_{i(qq)B}(x,y) + \Gamma_{i(qgq)B}(x,y)\ .
\label{combine}
\end{equation}
One may argue that this is not the only 
way to obtain gauge invariant correlations.  In particular, 
the combination $(y-x)\Gamma^\alpha_{i(qgq)B}(x,y)$ is the light-cone
expression for a gauge invariant distribution involving 
the gluon field strength rather than the covariant derivative.
On the other hand, these contributions necessarily involve
coefficients $C_{i(qgq)\alpha}^{\mu\nu}(x,y)$ that vanish
when $x=y$.  Eq.(\ref{vanish}) then implies that one can 
consider instead the combination of Eq.(\ref{combine})
free of charge.  Hence these are the only distributions 
relevant to our process.
Using hermitian conjugation, it is easy to show that $K_{1B}$ 
is symmetric in $x$ and $y$ and $K_{2B}$ is antisymmetric. 
We will use these symmetries to simplify our presentation of the 
coefficient functions.

\begin{figure}[t]
\begin{center}
\epsfig{file=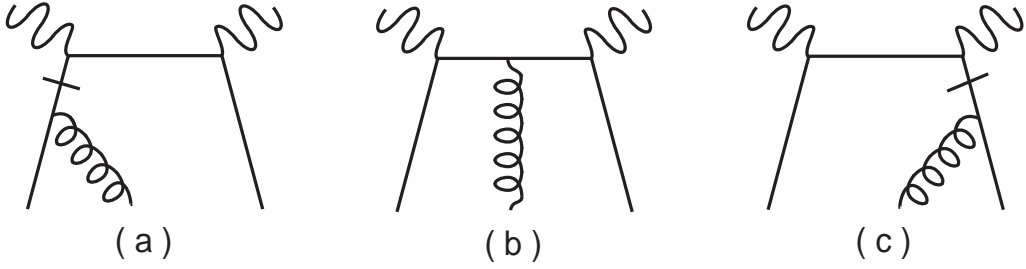,clip=,height=3.5cm,angle=0}
\end{center}
\caption{Tree contribution to $M^{\mu\nu}_{qgq}$. The contribution
to $M^{\mu\nu}_{qq}$ is obtained by replacing the gluon vertex
by $i\gamma^\alpha_\perp$.}
\label{fig:3}
\end{figure}

Let us consider the use of the above approach to the tree diagrams
shown in Fig. 3, where we have shown $M_{gqg}^{\mu\nu}$ only. Those
diagrams 
for $M_{qq}^{\mu\nu}$ can be obtained simply by replacing
the gluon interaction $it^a\gamma^\alpha_\perp$ by the 
transverse vector vertex $i\gamma^\alpha_\perp$. Figure 3
corresponds to the following on-shell process:
a quark and gluon with four-momenta $xp^\mu$ and 
$(y-x)p^\mu$, respectively, scatter with a photon of four-momentum 
$q^\mu$, producing a quark of four-momentum $yp^\mu$
and a forward photon. The Compton amplitude 
can easily be calculated using the usual Feynman rules.
When the quark and gluon combine before or 
after interacting with the photons, the intermediate propagator,
$i/(y\not\!p)$, drawn with a bar in Fig.~3, is singular and needs regularization. Adding
an infinitesimal $\lambda\not\! n$, one arrives at
\begin{equation} 
     {i(y\not\! p + \lambda \not\! n)\over (yp+\lambda n)^2}
     = {i \not\! p\over 2\lambda } +  {i \not\! n\over 2y} \ . 
\end{equation}
The first term yields zero when acting on the external
quark wave function, so only the second term contributes.
Combining the result from $M_{qq}^{\mu\nu}$ and taking into account
crossing symmetry, we find the following 
gauge-invariant expression for the tree-level Compton amplitude
\begin{eqnarray}
    T^{\mu\nu} 
  && = - i\epsilon^{\mu\nu\alpha\beta}q_\alpha S_{\perp\beta}
   {1\over \nu} \int dxdy \sum_i \hat e^2_i
  \nonumber \\
&& \times \;{2\over x(x_B-x)}\left[ K_{1i}(x,y) 
       + K_{2i}(x, y)\right]-(x_B\rightarrow -x_B)\ ,
\end{eqnarray}
where $\hat e^2_i=e^2_i-\bar e^2$ is the non-singlet 
part of the quark charge squared and $\bar e^2 = \sum_i e_i^2/n_f$.
The integration has support only for $\{x,y\}\in\lbrack -1,1\rbrack$.
When $|x_B|<1$, the above expression develops an imaginary part
through $x_B\rightarrow x_B-i\epsilon$. Using the optical 
theorem, one obtains the following tree result for $g^{NS}_T$, 
\begin{eqnarray}
    g_T^{NS(0)}(x_B, Q^2) = {1\over 2} \sum_i \hat e_i^2
   {2\over x_B}\int^1_{-1} dy\left(K_{1i}(x_B,y) 
   + K_{2i}(x_B,y)\right)  + (x_B\rightarrow -x_B)\ . 
\end{eqnarray}
>From QCD equations of motion, one can show
\begin{equation}
    \Delta q_T(x) = 
{2\over x}\int^1_{-1} dy\left(K_1(x, y) + K_2(x, y)\right)\ , 
\end{equation}
where $\Delta q_T(x)$ is defined as 
\begin{equation}
    \Delta q_T(x) = {1\over 2} \int {d\lambda\over 2\pi}
   e^{i\lambda x} \langle PS_\perp |\bar \psi(x)\gamma_\perp\gamma_5\psi(0) 
   |PS_\perp\rangle \ . 
\end{equation} 
This object seems to have a simple physical interpretation.

The tree level result suggests the following all-order
factorization formula for the non-singlet part of $g_T$, 
\begin{eqnarray}
    g^{NS}_T(x_B, Q^2) &=& \sum_i \hat e_i^2\int^1_{-1}
         { dxdy\over xy} \left(C_{1}\left({x_B\over x},
       {x_B\over y}, \alpha_s\right)K_{1i}(x,y)  \right. \nonumber \\
     && \left.+ C_{2}\left({x_B\over x},
       {x_B\over y}, \alpha_s\right)K_{2i}(x,y) \right)  +
     (x_B\rightarrow -x_B) \ .
\label{result}
\end{eqnarray}
$C_{1,2}$ can be written in terms of a perturbation
series in the strong coupling $\alpha_s$,
\begin{equation}
    C_{1,2} = \sum_i  C_{1,2}^{(i)}\left({\alpha_s\over2\pi}\right)^i \ . 
\end{equation}
The tree result yields
\begin{equation}
   C_{1,2}^{(0)}\left({x_B\over x},{x_B\over y}\right) = y\delta(x-x_B) \;\; .
  \end{equation}
 
\begin{figure}[p]
\begin{center}
\epsfig{file=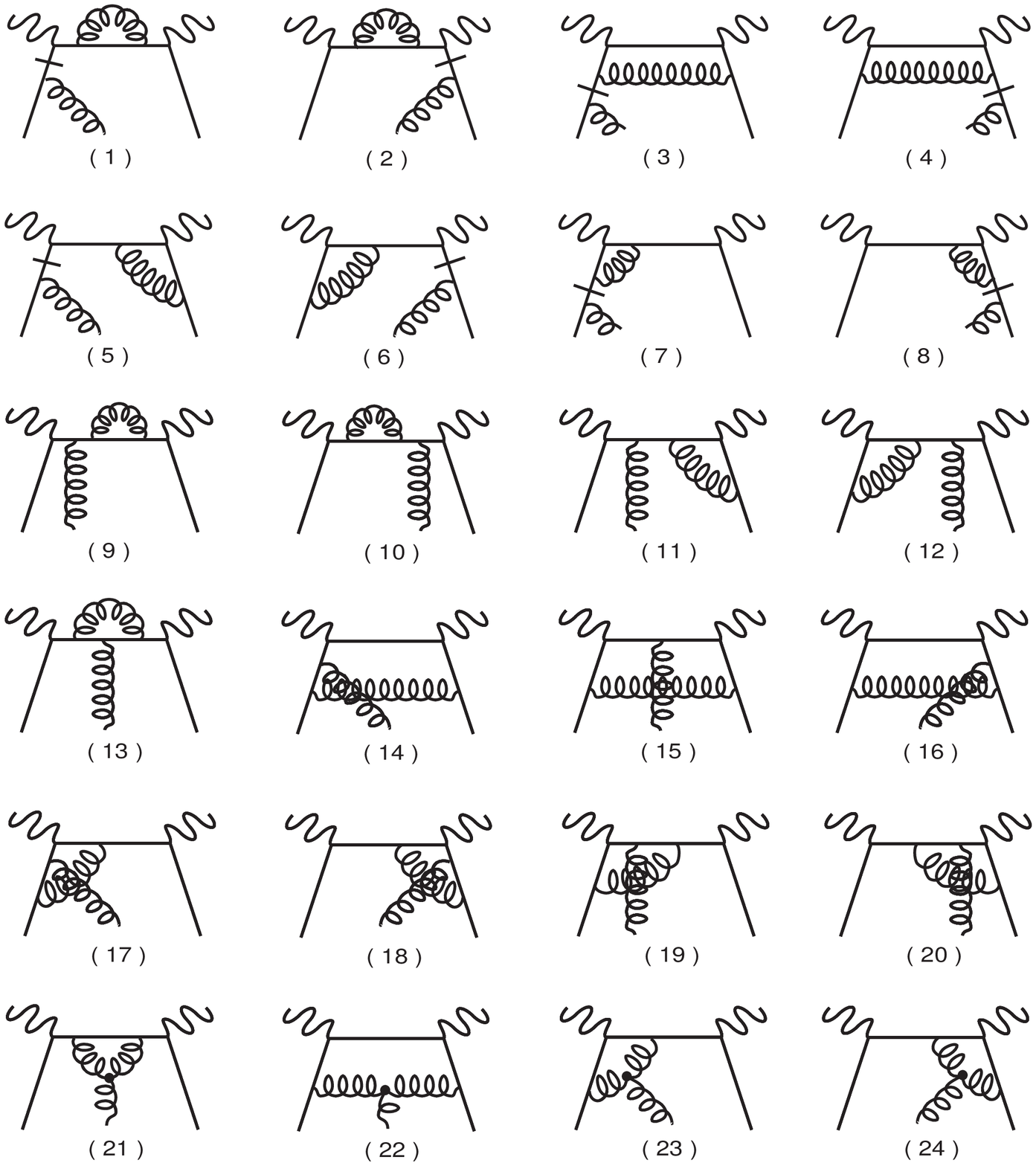,clip=,height=17cm,angle=0}
\end{center}
\caption{One-loop contribution to $M^{\mu\nu}_{qgq}$. The contribution
to $M^{\mu\nu}_{qq}$ is obtained by modifying the external gluon vertex
in the first 20 diagrams.} 
\label{fig:4}
\end{figure}   

Now we come to the main subject of the paper: one-loop 
radiative corrections for $g^{NS}_T(x_B,Q^2)$. All the Feynman diagrams
for $M_{qgq}^{\mu\nu}$ are shown in Fig. 4. The first 12 diagrams 
have color factor $C_F$, the next 8 come with the 
factor $C_F-C_A/2$, 
and the final 4 with $C_A=N_c$. Since the external transverse 
momentum of the quark only goes through the quark propagators, the
Taylor-expansion of $M_{qq}^{\mu\nu}$ yields the vector vertex 
insertion on the quark line only. These diagrams correspond to the 
first 20 diagrams in Fig. 4 with the external gluon vertex replaced 
by the vector vertex. An inspection of the one-loop Feynman integrals
reveals a simple rule: $M_{qq}^{\mu\nu}$ is just the $C_F$ part
of $M_{qgq}^{\mu\nu}$ and, hence, the former need not be 
calculated separately. 

Since all Feynman diagrams are computed in bare perturbation theory, 
the result depends on the bare coupling $g_B$. We replace
it with a renormalized coupling in the $\overline{\rm MS}$ 
scheme, with a difference of higher order in $\alpha_s(\mu^2)$. 
After a lengthy calculation involving one-loop integrals with up to five 
internal Feynman propagators, we
find 
\begin{eqnarray}
   S_T^{(1)NS}(x_B, Q^2) 
  &=& \sum_i \hat e_i^2 \int {dxdy\over xy}
   \left[M^{(1)}_{1}\left({x\over x_B}, {y\over x_B}\right)K_{1iB}(x, y)  
\right.\nonumber\\ &&\left.+ 
   M^{(1)}_{2}\left({x\over x_B}, {y\over x_B}\right)K_{2iB}(x, y) 
  \right]
   - (x_B\rightarrow - x_B)\ . 
\end{eqnarray}    
Explicit expressions for the $M$'s in the region 
$|x_B|>1$ can be found in the
Appendix.

Now we are ready to show that $S^{NS}_T(x_B, Q^2)$ is 
factorizable at the one-loop level, i.e., the infrared 
poles $1/\epsilon$ in $M^{(1)}_i$ match the ultraviolet 
poles in $K_{iB}$. To this end, we use the infrared poles in $M_i$ 
to generate a scale evolution equation for the parton
distributions. From the result in the Appendix, we find  
\begin{eqnarray}
 {d\over d \ln \mu^2}\int^1_{-1} &dxdy&\left({2\over
    x(x_B-x)}\right)(K_1(x,y,\mu^2) + K_2(x,y,\mu^2))
  \nonumber \\
  &=& -{\alpha_s(\mu^2)\over 2\pi}
  \int^1_{-1} dxdy\left\{C_FK_1(x,y,\mu^2)
  \left[{-1\over x(x_B-x)} + {2\over x^2}
     - {2\over xy} \right.\right. \nonumber \\
 && - 2\left({2x_B\over x^2y} +{1\over x^2} - {x_B\over x^3}
            + {2\over x(x_B-x)}\right)
 \log \left({x_B-x\over x_B}\right)  
  \nonumber \\
&& \left.
     + {2\over y-x}\left({2x_B\over x(y-x)} + {1\over x_B-y}
   \right)\log \left({x_B-y+x\over x_B}\right)\right]
   \nonumber \\
  &&  + C_F K_2(x, y, \mu^2) 
\left[{-1\over x(x_B-x)} + {2\over x^2}
      - 2\left({1\over x^2} - {x_B\over x^3}
            + {2\over x(x_B-x)}\right)
\right.  \nonumber \\
     && \left.
 \times \log\left( {x_B-x\over x_B}\right)  
     - {2\over (y-x)(x_B-y)}\log \left({x_B-y+x\over x_B}
  \right)\right]
   \nonumber \\
  &&  + {C_A\over 2}K_1(x,y,\mu^2)
\left[
      - {4x_B \over xy(y-x)}\log \left({x_B-x\over x_B}\right)  
\right. \nonumber \\
     && \left.
     - {2x_B\over y-x} \left({2\over x(y-x)} + {1\over x_B-y}\right)
\log\left({x_B-y+x\over x_B}\right)\right]
  \nonumber \\ && 
   \left.
   + {C_A\over 2}K_2(x,y,\mu^2)
  {2\over (y-x)(x_B-y)}\log\left({x_B-y+x\over x_B}\right)
\right\} \ . 
\end{eqnarray}
Expanding the above in the large $x_B$ limit,
we get the evolution equation for the moments of the parton
distributions.  A detailed check shows that the equation 
is identical to that in Ref. \cite{bkl} obtained by 
studying the ultraviolet divergences present in the
twist-three operators. In particular, in the large $N_c$
limit, the above equation becomes autonomous \cite{ali,osborne}.

The final step of the calculation is to take the imaginary part 
of the factorized $S^{NS}_T(x_B, Q^2)$ to get a factorized expression for
the structure function $g^{NS}_T(x_B, Q^2)$
in the physical region $x_B<1$. With the definition in Eq. (19)
we find the following result for the coefficient functions, 
\begin{eqnarray}
   C_1^{(1)}\left({x_B\over x}, {x_B\over y}\right) &&= 
  C_F y\left\{-2\delta(x-x_B) + \theta\left({x\over x_B}-1\right)
 \left[{3/2\over (x_B-x)_+} - {x_B\over x^2} + {6x_B\over xy}
  \right.\right. \nonumber \\ 
  && + {2x_B\over y(y-x)}  - {1/2\over x} - {1\over y}
  -{1\over y-x} 
   - \left({2x_B\over xy}
  + {1\over x} - {x_B\over x^2}\right)
  \log\left({x-x_B\over x_B}\right) \nonumber \\
 && \left. -\left({2\over (x_B-x)}
   \log\left({x-x_B\over x_B}\right)\right)_+\right] \nonumber \\
 && + \theta\left({y-x\over x_B}-1\right)
   {1\over y-x}\left[1
  - {2x\over x_B-y} - {4x_B\over y-x} \right. \nonumber \\
  &&   \left. + \left({2x_B\over y-x} + {x\over x_B-y}
  \right)\log{\left(y-x-x_B\over x_B\right)}\right]
\nonumber \\ &&
   + \delta(x-x_B)\log{\left(1-x_B\over x_B\right)}
     \left[-{3\over 2} + \log{\left(1-x_B\over x_B\right)}\right]
\nonumber \\ &&
 \left.  + \delta(y-x_B){2x\over x_B-x} \log\left({x\over x_B}\right)
    \left[1-{1\over 4}\log\left({x\over x_B}\right)\right]
\right\} \nonumber\\
&&+
 {C_A\over 2}y \left\{ \theta\left({x\over x_B}-1\right){x_B\over y}
\left[ {1\over x_B-y} + {4\over y-x}\left(1-{1\over 2} 
   \log\left({x_B-x\over x_B}\right)\right)\right] 
\right.  \nonumber \\
 &&  -\theta\left({y-x\over x_B}-1\right){1\over (y-x)}\left[
   1 - \left({2x\over x_B-y} + {4x_B\over (y-x)}\right) \right.
  \nonumber \\ &&
 \left. \times \left(1-{1\over 2}
   \log\left({y-x-x_B\over x_B}\right)\right)\right] 
 - \delta(y-x_B)\left[ \log\left({x-x_B\over x_B}\right) \right.
  \nonumber \\
   && \left.\left.  + {2x\over (x_B-x)}
  \left(1-{1\over 4}\log\left({x\over x_B}\right)
  \right)\log\left({x\over x_B}\right)\right] \right\}  \ ,   
\end{eqnarray}
\begin{eqnarray}
  C_2^{(1)}\left({x_B\over x},{x_B\over y}\right)
&&= C_F y\left\{-2\delta(x-x_B) + \theta\left({x\over x_B}-1\right)
 \left[{3/2\over (x_B-x)_+} - {x_B\over x^2} 
  \right.\right. \nonumber \\ 
  &&  - {1/2\over x} + {1\over y} 
   \left.- \left(
  {1\over x} - {x_B\over x^2}\right)\log\left(
  {x-x_B\over x_B}\right) - \left({2\over (x_B-x)}
   \log\left({x-x_B\over x_B}\right)\right)_+\right] \nonumber \\
 && + \theta\left({y-x\over x_B}-1\right)
   {1\over y-x}\left[1
  + {2x\over x_B-y} - {x\over x_B-y}\log{\left(y-x-x_B\over x_B\right)}\right]
\nonumber \\ &&
   + \delta(x-x_B)\log{\left(1-x_B\over x_B\right)}
     \left[-{3\over 2} + \log{\left(1-x_B\over x_B\right)}\right]
\nonumber \\ &&
 \left.  - \delta(y-x_B){2x\over x_B-x} \log\left({x\over x_B}\right)
    \left[1-{1\over 4}\log\left({x\over x_B}\right)\right]
\right\} 
 \nonumber \\
 && +  
  {C_A\over 2}y \left\{ -{x_B\over y(x_B-y)} \theta\left({x\over x_B}
-1\right)
  \right. \nonumber \\
 && 
   - {1\over y-x}\left[1 + {2x\over x_B-y} \left(
  1- {1\over 2}\log\left({y-x-x_B\over x_B}\right)\right)\right]
\theta\left({y-x\over x_B}-1\right) \nonumber \\
  && \left. + \delta(y-x_B)
   \left[ \log\left({x-x_B\over x_B}\right)
   + {2x\over x_B-x}\left(1-{1\over 4}\log \left({x\over x_B}\right)
   \right)\log\left({x\over x_B}\right)\right] \right\} 
\end{eqnarray}
The definition of the + functions can be found in 
Ref. \cite{dglap}. The support for the parton
correlations limits $y-x$ to the interval $[-1,1]$. 

To check the Burkhardt-Cottingham sum rule \cite{bc}, we integrate 
$g^{NS}_T(x_B,Q^2)$ over $x_B$.  Assuming the integration over 
$x$ and $y$ can be interchanged with that of 
$x_B$, one obtains
\begin{equation}
  \int^1_0 dx_B g^{NS}_T(x_B, Q^2)  = {1\over 2}\sum_i \hat e_i^2
   \left(1-{7\over 2}C_F{\alpha_s\over 2\pi}\right)
   \langle PS|\overline{\psi}_i\gamma_\perp\gamma_5\psi_i|PS\rangle \ . 
\end{equation}
Here the coefficient $7/2$ reduces to $3/2$ if we define $\gamma_5$
so that the non-singlet axial current is conserved. Compared
with the factorization formula for $g_1(x_B, Q^2)$, we
have the Burkhardt-Cottingham sum rule at one loop
\begin{equation}
   \int^1_0 dx_B g^{NS}_2(x_B, Q^2) = 0 \ . 
\end{equation}
If the order of integration cannot be interchanged 
because of the singular behevior of the parton distributions at small 
$x$ and $y$, the above sum rule may be violated. Indeed
some small $x_B$ study does indicate such singular behavior
\cite{smallx}. 

Finally, we consider the next-to-leading order correction
to the non-singlet part of the $x^2$ moment of 
$g_T(x_B, Q^2)$. In the leading order, it is well known :
\begin{equation}
     \int^1_0 dx~x^2g^{NS}_T(x, Q^2) 
   = {1\over 3} \sum_i \hat e_i^2
   \left({1\over 2}a_{2i}(Q^2) + d_{2i}(Q^2)\right) \ ,
\end{equation}
where $a_2$ is the second moment of the $g_1(x,Q^2)$ structure
function and $d_2$ is a twist-three matrix element \cite{jj}. 
Using the coefficient functions found above, we see that
\begin{eqnarray}
    \int^1_0 dx x^2 g^{NS}_T(x,Q^2) 
  &=& {1\over 3}\sum_i\hat e_i^2
    \left\lbrace{a_{2i}\over 2}\left\lbrack 1+{\alpha_s(Q^2)\over 4\pi}
  {7\over 12}C_F\right\rbrack \right. \nonumber \\
 && \phantom{{1\over 3}\sum_i\hat e_i^2}\quad
\left. + d_{2i}\left\lbrack1+{\alpha_s(Q^2)\over 4\pi}\left({27\over 4}C_A
  - {29\over 3}C_F\right)\right\rbrack\right\rbrace\ . 
\end{eqnarray}
Using the next-to-leading result for $g_1(x, Q^2)$\cite{kodaira}, 
we find 
\begin{equation}
   \int^1_0 dx x^2\left(g^{NS}_T(x, Q^2)-{1\over 3}g^{NS}_1(x, Q^2)\right)  
  = {1\over 3}\sum_i \hat e_i^2
 d_{2i}\left(1+{\alpha_s(Q^2)\over 4\pi}\left({27\over 4}C_A
  - {29\over 3}C_F\right)\right)  \ . 
\end{equation}
Notice that the combination of $g_T$ and $g_1$ 
relevant to $d_2$ receives
no radiative correction.

To summarize, we have extended the leading-twist
factorization formalism to higher twist. The extension 
mainly involves a correct identification of the 
intermediate parton states and arranging different calculations 
into gauge invariant combinations. Because of the multipartons in
the initial state, higher-twist perturbative calculations are, 
in general, much more complicated than the leading twist cases.
As an example, we have presented the one-loop coefficient
functions the twist-three process involving forward
Compton scattering on a transversely polarized nucleon target.
This result can be combined with a two-loop calculation
of the twist-three evolution equation to yield a next-to-leading
order formalism for the $g_2$ structure function. 

\acknowledgements
The authors wish to acknowledge the support of the U.S.~Department
of Energy under grant no. DE-FG02-93ER-40762.
X. Song was supported in part by the Institute of Nuclear and
Particle Physics, Department of Physics, University of Virginia, and
the Commonwealth of Virginia. 
\appendix
\section*{Appendix}
In this appendix, we present the tree and one-loop 
Compton amplitudes for virtual photon scattering 
on a transversely polarized nucleon in the unfactorized form. 
To simplify the expressions, we assume $|x_B| >1$ so that
the amplitudes are purely real. 
With $\epsilon^{0123}=+1$, 
we write the Compton tensor as
\begin{equation}
    T^{\mu\nu} = 
  -i\epsilon^{\mu\nu\alpha\beta}q_\alpha S_{\perp\beta}{1\over 
  \nu} S_T(x_B, Q^2) \ . 
\end{equation}
In QCD, we can write,
\begin{eqnarray}
   S^{NS}_T(x_B, Q^2) 
  &=& \sum_i \hat e_i^2 \int^1_{-1} {dxdy\over xy}
   \left[M_{1}\left({x\over x_B}, {y\over x_B}\right)K_{1iB}(x, y) \right. 
\nonumber\\&&\left.+ 
   M_{2}\left({x\over x_B}, {y\over x_B}\right)K_{2iB}(x, y) \right]
   - (x_B\rightarrow - x_B) \ , 
\end{eqnarray}    
where the $K$'s are unrenormalized parton distributions as 
defined in the text, $M$'s are perturbation series in $\alpha_s$ and
have infrared poles.  

At tree level, the result is well known,
\begin{equation}   
     M^{(0)}_1(x, y) = M^{(0)}_2(x,y) = {2y\over 1-x} \ .
   \end{equation}
At the one-loop level, the following amplitudes are associated with 
distributions with a covariant derivative, 
\begin{eqnarray}  
    M^{(1)}_1(x, y) &=& {\alpha_{sB}\over 2\pi}C_F
  \left({\bar\mu^2\over Q^2}\right)^{\epsilon/2} \left\{
       {2\over \epsilon} \left[{-y\over (1-x)} + {2y\over x}
     - 2 - 2\left({2\over x} +{y\over x} - {y\over x^2}
            + {2y\over 1-x}\right)
\right.\right.\nonumber \\
     && \left.
 \times \log (1-x)  
     + {2y\over y-x}\left({2\over y-x} + {x\over 1-y}
   \right)\log (1-y+x)\right]
  \nonumber \\ && 
  -2\left({2y\over 1-x} - {y\over x} + 3\right)
   -\left({3y\over 1-x} - {2y\over x^2} \right. 
  \nonumber \\ &&
   +  \left. {12\over x} + {4\over y-x} - {y\over x}
   -2-{2y\over y-x}\right)\log(1-x)
   \nonumber \\ &&
   -{2y\over y-x}\left(1 -{2x\over 1-y}
    -{4\over y-x}\right) \log(1-y+x)
   \nonumber \\ &&
  +\left({2\over x} + {y\over x} - {y\over x^2}
     + {2y\over 1-x} \right) \log^2(1-x)
  \nonumber \\ && \left.
  - {y\over y-x}\left({2\over y-x} + {x\over 1-y}\right)
    \log^2(1-y+x)
\right\} \nonumber\\
&&+{\alpha_{sB}\over 2\pi}{C_A\over 2}
  \left({\bar\mu^2\over Q^2}\right)^{\epsilon/2} \left\{
       {2\over \epsilon} \left[
      - {4 \over y-x}\log (1-x)  
\right.\right.\nonumber \\
     && \left.
     - {2y\over y-x} \left({2\over y-x} + {x\over 1-y}\right)\log (1-y+x)\right]
  \nonumber \\ && 
   -2\left({1\over 1-y} +{4\over y-x} \right)\log(1-x)
   \nonumber \\ &&
   +{2y\over y-x}\left(1 -{2x\over 1-y}- {4\over y-x}\right) \log(1-y+x)
   \nonumber \\ &&
  + {2\over y-x}\log^2(1-x)
  \nonumber \\ && \left.
  + {y\over y-x} \left({2\over y-x} + {x\over 1-y}\right)
    \log^2(1-y+x)
\right\}\ ,
\end{eqnarray}

\begin{eqnarray}  
    M^{(1)}_2(x, y) &=& {\alpha_{sB}\over 2\pi}C_F
  \left({\bar\mu^2\over Q^2}\right)^{\epsilon/2} \left\{
       {2\over \epsilon} \left[{-y\over (1-x)} + {2y\over x}
      - 2\left({y\over x} - {y\over x^2}
            + {2y\over 1-x}\right)
\right.\right.\nonumber \\
     && \left.
 \times \log (1-x)  
     - {2xy\over (y-x)(1-y)}\log (1-y+x)\right]
  \nonumber \\ && 
  -2\left({2y\over 1-x} - {y\over x}\right)
   -\left({3y\over 1-x} - {2y\over x^2} - {y\over x}+2\right)\log(1-x)
   \nonumber \\ &&
   -{2y\over y-x}\left(1 +{2x\over 1-y}\right) \log(1-y+x)
   \nonumber \\ &&
  +\left({y\over x} - {y\over x^2}
     + {2y\over 1-x} \right) \log^2(1-x)
  \nonumber \\ && \left.
  + {xy\over (y-x)(1-y)}
    \log^2(1-y+x)
\right\} \nonumber \\ &&
   + {\alpha_{sB}\over 2\pi}{C_A\over 2}
  \left({\bar\mu^2\over Q^2}\right)^{\epsilon/2} \left\{
       {2\over \epsilon} 
     {2xy\over (y-x)(1-y)}\log (1-y+x)\right.
  \nonumber \\ && 
   + {2\over 1-y} \log(1-x)
   \nonumber \\ &&
   +{2y\over y-x}\left(1 +{2x\over 1-y}\right) \log(1-y+x)
  \nonumber \\ && \left.
  - {xy\over (y-x)(1-y)}\log^2(1-y+x)
\right\}\ , 
\end{eqnarray}
where $\epsilon=4-d$, $\bar\mu^2 = 4\pi e^{-\gamma_E}\mu^2$, and 
$\gamma_E$ is the Euler constant.


\begin{thebibliography}{}
\bibitem{mueller}
A. Mueller, {\it Perturbative Quantum Chromodynamics} (World 
Scientific, Singapore, 1989). 

\bibitem{dglap}
V. N. Gribov and L. N. Lipatov, Sov. J. Nucl. Phys. {\bf 15}, 78 (1972); \\
G. Altarelli and G. Parisi, Nucl. Phys. B {\bf 126}, 298 (1977); \\
Y. L. Dokshitser, Sov. Phys.-JETP {\bf 46}, 641 (1977). 

\bibitem{sterman}
J. Qiu and G. Sterman, Nucl. Phys. B {\bf 353}, 137 (1991). 

\bibitem{e155x}
P. Bosted for E155x Collaboration, Nucl. Phys. A {\bf 666-667}, 300 (2000).

\bibitem{jlab}
JLab workshop on 12 GeV upgrade, January, 2000.   

\bibitem{hey}
A. J. G. Hey and J. E. Mandula, Phys. Rev. D {\bf 5}, 2610 (1972);\\
M. A. Ahmed and G. G. Ross, Nucl. Phys. B {\bf 111}, 441 (1976); \\
K. Sasaki, Prog. Theor. Phys. {\bf 54}, 1816 (1975). 

\bibitem{kodaira}
J. Kodaira, S. Matsuda, K. Sasaki, T. Uematsu, Nucl. Phys. B 
{\bf 159}, 99 (1979). 

\bibitem{quarkonly}
R. Mertig and W. L. van Neervan, Z. Phys. C {\bf 60}, 489 (1993); \\
G. Altarelli, B. Lampe, P. Nason and G. Ridolfi, 
Phys. Lett. B {\bf 334}, 187 (1994); \\
J. Kodaira, S. Matsuda, T. Uematsu, and K. Sasaki, 
Phys. Lett. B {\bf 345}, 527 (1995); \\
P. Mathews, V. Ravindran, and K. Sridhar, 
hep-ph/9607385; \\
A. Gabieli, G. Ridolfi Phys, Lett. B {\bf 417}, 369 (1998). 

\bibitem{sv}
E. V. Shuryak and A. I. Vainshtein, Nucl. Phys. {\bf B199}, 
951 (1982). 

\bibitem{bkl}
A. P. Bukhvostov, E. A. Kuraev, and L. N. Lipatov, 
JETP Lett. {\bf 37}, 484 (1983); 
Sov. Phys. JETP {\bf 60}, 22 (1984). 

\bibitem{other}
P. G. Ratcliffe, Nucl. Phys. B {\bf 264}, 493 (1989); \\
I. I. Balitsky and V. M. Braun, Nucl. Phys. B {\bf 311}, 541
(1989); \\
X. Ji and C. Chou, Phys. Rev. D {\bf 42}, 3637 (1990);\\
B. Geyer, D. M\"uller and D. Robaschik, hep-ph/9611452;\\
J. Kodaira, Y. Yasui, K. Tanaka and T. Uematsu, Phys. Lett. B387, 855 (1996).

\bibitem{et}
A. V. Efremov and O. V. Teryaev, Sov. J. Nucl. Phys. {\bf 36}, 140 (1982). 

\bibitem{jj}
R. L. Jaffe and X. Ji, Phys. Rev. D {\bf 43}, 724 (1991). 

\bibitem{ji}
X. Ji, Nucl. Phys. B {\bf 402}, 217 (1993). 

\bibitem{efp}
G. Curci, W. Furmanski, and R. Petronzio, Nucl. Phys. B {\bf 175}, 27 (1980). 

\bibitem{qs}
J. Qiu and G. Sterman, Phys. Rev. Lett. {\bf 67}, 2264 (1991); 
Phys. Rev. D {\bf 59}, 014004 (1999). 

\bibitem{ali}
A. Ali, V. M. Braun and G. Hiller, Phys. Lett. 
B {\bf 266}, 117 (1991).

\bibitem{osborne}
X. Ji and J. Osborne, Eur. Phys. J. C {\bf 9}, 487 (1999). 

\bibitem{bc}
H. Burkhardt and W. N. Cottingham, Ann. Phys. (N. Y.) {\bf 56}, 453 (1970).  

\bibitem{smallx}
I. P. Ivanov, N. N. Nikolaev, A. V. Pronyaev, W. Schafer, 
Phys. Lett. B {\bf 457}, 218 (1999). 

\end{thebibliography}
\end{document}